\documentclass[aps,prl,twocolumn,showpacs,preprintnumbers,amsmath,amssymb,superscriptaddress]{revtex4}
\usepackage{amsmath}
\usepackage{amssymb}

\usepackage{epsfig}
\usepackage{dcolumn}
\usepackage{bm}
\usepackage{color} 

\begin{document}

\preprint{To submit for PRL or CPC or ...}

\title{Pure Monte Carlo Method: a Third Way for Plasma Simulation}
\author{Hua-sheng XIE}
\affiliation{Institute for Fusion Theory and Simulation, Zhejiang
University, Hangzhou 310027, People's Republic of China}
\date{\today}

\begin{abstract}
We bring a totally new concept for plasma simulation, other than the
conventional two ways: Fluid/Kinetic Continuum (FKC) method and
Particle-in-Cell (PIC) method. This method is based on Pure Monte
Carlo (PMC), but far beyond traditional treatments. PMC solves all
the equations (kinetic, fluid, field) and treats all the procedures
(collisions, others) in the system via MC method. As shown in two
paradigms, many advantages have found. It has shown the capability
to be the third importance approach for plasma simulation or even
completely substitute the other two in the future. It's also
suitable for many unsolved problems, then bring plasma simulation to
a new era.
\end{abstract} 
\maketitle \textbf{\textit{Introduction}}--The original of Monte
Carlo (MC) method for computations can back to hundreds of years
ago, e.g., the Buffon's needle experiment. With the emergence of the
modern computer in 1940s and pioneer works by John von Neumann,
Stanislaw Ulam and Nicholas Metropolis, this method brings amounts
of practical applications. The famous and classical [Metropolis1953]
paper is said \cite{Gubernatis2005} mainly contributed by the hero
of plasma physics, M. N. Rosenbluth. However, up to now, MC for
physics is mainly on statistical physics, and only in merely special
cases for plasma physics.

Some people had known that, in fact, the PIC method can be seen a
type of MC method. \cite{Aydemir1994} provides a unified MC
interpretation of particle simulations, which then can use the MC
theory to estimate the errors and develop many types of sampling
approaches, e.g., simple MC, importance sampling, the $\delta$-f
method and more.

While, at most, the conventionally PIC only treats the kinetic
equation using MC viewpoint and the field equations are still solved
by Deterministic Discretizations (DD, finite differences and so on).
FKC is solved totally by DD. Generally, each approach has its
advantages and disadvantages. The main disadvantages of DD are the
numerical unstability and dissipation and not easy to treat complex
geometries and complex procedures which cannot be described easily
by differential equations (ODEs, PDEs). To avoid old disadvantages,
a way out is to develop new approaches.

We find, indeed, we can have a third way for plasma simulation,
i.e., we use totally stochastic statistical (Monte Carlo) approach.
We need only a change of concept. In Copenhagen interpretation of
quantum mechanics, we have known that the intrinsic of the world is
uncertainty and all things are probabilities, only the statistic is
determined. If we can accept this, we can easily accept the
philosophy of this PMC letter. The remaining thing is how we do
that, or exactly, how we solve differential equations (in fluid and
kinetic) via MC. A branch of mathematic called Stochastic
Differential Equations (SDE) has provided us some basic tools.

At below, we will firstly describe the PMC method and then treat a
fluid edge pedestal transport model and the kinetic 1D electrostatic
problem as paradigms.

\textbf{\textit{PMC Method}}--In plasma simulation, MC has been used
to many complex procedures, such as collisions and impurities
transport, which has been widely known. So, here, we need not talk
too much about that, and all the old MC methods in plasma simulation
can be inherited by PMC method naturally. The new steps of PMC are
to tell how we solve the fluid and field equations.

Using MC to solve PDEs can back to two pivotal papers,
\cite{Courant1928} and \cite{Metropolis1949}, the former uses
probabilistic interpretations for linear elliptic and parabolic
equations, the latter gives MC methods for solving
integro-differential equations occur in various branches of the
natural sciences.

\textit{Feynman-Kac formula}--Many types of plasma equations (e.g.,
transport equations, Vlasov equation) have the form:
\begin{equation}\label{eqn1}
{{\partial f} \over {\partial t}}+\mu (x,t){{\partial f} \over
{\partial x}} - {1 \over 2}{\sigma ^2}(x,t){{{\partial ^2}f} \over
{\partial {x^2}}} - V(x,t)f + p(x,t)=0.
\end{equation}
Feynman-Kac formula \cite{Feynman1948}\cite{Kac1949} tells us,
Eq.(\ref{eqn1}) is equivalent to SDE when $V, p=0$ ($V, p{\neq}0$
case can also find),
\begin{equation}\label{eqn2}
dX = \mu (X,t)dt + \sigma (X,t)d{W_t},
\end{equation}
where $W_t$  is Wiener process (Brownian motion), which is described
by normal (Gaussian) distribution. Eqs.(\ref{eqn1}) and (\ref{eqn2})
are easily general to high dimensions or more variables, e.g.,
$(\mathbf{x},\mathbf{v},t)$. The above descriptions are also shown
by It\=o's lemma \cite{Itoh1951}, a classical work in SDE.

Now, we solve the convection-diffusion equation as example to show
how to use this formula, which is standard steps and well known by
SDE people but may not familiar by plasma community. At this case,
$a=\mu$ and $D=\sigma^2/2$, are constants. For $f(x,0)=f_0(x)$ , we
start with a set $N$ samples $\xi_1^0,\cdots,\xi_N^0$ from $f_0(x)$.
A sample for every next step is
\begin{equation}\label{eqn3}
\xi _i^{t+dt} = \xi _i^t + adt + \sqrt {2Ddt} {\eta _i},~i =
1,\cdots,N.
\end{equation}
where $\eta _i$  is normal distribution with zero mean and unit
variance. For $a=0$, we get pure diffusion, and for $D=0$, pure
convection. In fact, integral form explicit exact solution is
\begin{gather}\label{eqn4}
f(x,t) = \int {{G_t}(x-y){f_0}}(y-at)dy = {G_t} *
{f_0}(x-at), \\
 {G_t}(x-y) = {1 \over {{{( {4\pi Dt} )}^{1{\rm{/}}2}}}}{e^{ -
{{{{\left( {x - y} \right)}^2}} \over {4Dt}}}}. \notag
\end{gather}
If we use MC to integral (\ref{eqn4}) directly, and use $N'$ samples
for particles and $M$ points in space, the total cost will be
$O(N'\times M)$ . While using (\ref{eqn3}), the total cost is
$O(N)$. We write (\ref{eqn4}) here for benchmark.

\begin{figure}[htp]
\epsfig{file=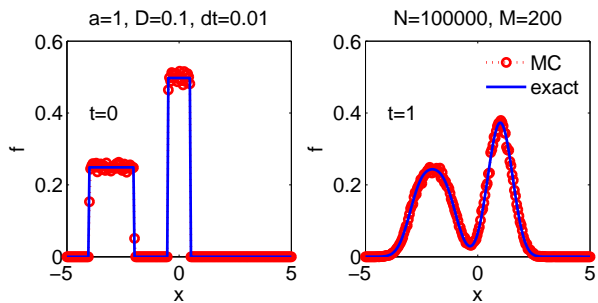,clip=true,width=9cm}
\caption{\label{fig1}(Color online) Solve convection-diffusion
equation using (\ref{eqn3}), as example to show PMC method.}
\end{figure}

Using (\ref{eqn3}), a result is show in FIG.\ref{fig1}. We can see
that the MC result reproduces the exact solution very well, when we
using larger $N$ and $M$, the result can be even better. Here, $M$
is not for calculation but just for reconstructing the grids for
$f$.

The above is a summary of SDE for our usage, which is a key
fundament of our PMC method, the applications to plasma examples
will be shown at below.

\begin{figure}[htp]
\epsfig{file=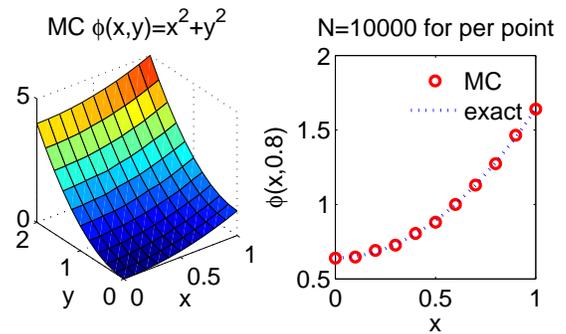,clip=true,width=9cm}
\caption{\label{fig2}(Color online) MC for 2D Poisson equation,
example.}
\end{figure}

\textit{Poisson equation}--We may meet Poisson equation frequently
in plasma physics, for example, in the electrostatic kinetic case or
the electromagnetic gyrokinetic case. The equation is as the form
\begin{equation}\label{eqn5}
{\nabla ^2}\phi (\mathbf{x}) =  - \rho (\mathbf{x}),~\mathbf{x} \in
D.
\end{equation}
If $\rho =0$, gives Laplace equation, which can be found solved by
MC in many places, e.g., \cite{Landau2008}. A good introduction of
MC methods for Poisson equation can find in \cite{Delaurentis1990}.
The method can also take from an extended Feynman-Kac formula. For
Dirichlet boundary, $\phi(\mathbf{x})=f(\mathbf{x}),~\mathbf{x} \in
\partial{D}$. Standard solution for position $\mathbf{x}_0$ using SDE
notation is
\begin{equation}\label{eqn6}
\phi \left(\mathbf{x}_0 \right) = {1 \over 2}E\left[ {\int_0^{{\tau
_{\partial D}}} {\rho \left( {{W_t}} \right)dt} } \right] + E\left[
{f\left( {{W_{{\tau _{\partial D}}}}} \right)} \right],
\end{equation}
where, $\tau_{\partial D}=inf\{t: W_t \in {\partial D} \}$  is the
first-passage time and $W_{\tau_{\partial D}}$ is the first-passage
location on the boundary, $E$ is average. Using floating random
walk, the results for 2D are show in FIG.\ref{fig2}.
The MC method here has another advantage that we can obtain the
solution at a few points directly instead of calculating the whole
range, and even if there are steep gradients or irregular boundary.
Some updates of MC for Poisson equation may also find in recent
literatures, such as for Neumann boundary, improving efficiencies,
or reducing errors.

\textit{Other equations}--There are many other types of PDEs in
plasma physics. Scalar conservation law equation
$\partial{u}/\partial{t}+\partial{F(u)}/\partial{x}=0$, e.g.,
Burgers' equation, can find be solved generally in \cite{Jin1995}.
Navier-Stokes (vector) equation for turbulence can find in
\cite{Belotserkovskii2010}. For ODE, we often need just the standard
MC integral. Some more MC methods and details for electromagnetic
problems are introduced in \cite{Sadiku2009}. One can also refer
\cite{Kroese2011} for some general descriptions.

The types of equations mentioned at above have contented many of the
main plasma physics equations.

\textbf{\textit{Application Paradigms}}--We use the above methods
 to solve two practical plasma problems.

\textit{Simple fluid edge pedestal transport model}--The simplest
model for calculating pedestal width can find in \cite{Stangeby2000}
or \cite{Wesson2004}, the inner range of SOL (Scrape-Off Layer, for
divertor or limiter) is mainly slowly perpendicular transport, and
the outer range is mainly fast parallel (to the magnetic field line,
mainly on toroidal $\varphi$  direction) transport and the particles
travel at most  $L$ ($2{\pi}R$ or $2{\pi}qR$) then hit the target
with the thermal velocity, where $R$  is the major radius, $q$  is
edge safety factor, and the thermal velocity is equal to acoustic
velocity $c_s$ at edge from sheath calculations.

We can write the inner range transport equation as
\begin{equation}\label{eqn7}
{{\partial n} \over {\partial t}} = {1 \over r}{\partial \over
{\partial r}}\left( {rD{{\partial n} \over {\partial r}}} \right) +
S(n,r,t) = {D \over r}{{\partial n} \over {\partial r}} +
D{{{\partial ^2}n} \over {\partial {r^2}}} + S.
\end{equation}
For simplification, we only use the density diffusion equation and
treat the coefficient $D$ as constant, and also ignore the source
term $S$. For outer range, we find quickly that we cannot easily
write a 1D equation combine to the inner range equation. Even though
we write down the outer range equation, we will quickly meet the
singularity problem at the last closed (magnetic) flux surface
(LCFS, separatrix). A special treatment for transport barriers can
find in \cite{Tokar2006}, which rewrites the transport equations to
new form. However, using PMC, we can avoid the above problems
automatically.

The inner equation (\ref{eqn7}) can be solved using (\ref{eqn2}).
The outer range needs an extra MC treatment: if a particle comes out
to the outer range, we use the same equation as the inner range for
perpendicular transport but in the same time let it vanish random in
time $[0, T)$ if it is still in the outer range, with $T=L/c_s$.

Quickly, we can find when using (\ref{eqn3}) to treat
(\ref{eqn7}), the drift velocity $a=-D/r \to \infty$ for the MC
particles at $r \to 0$. This problem is also met similarly in PIC
\cite{Birdsall1991} when using polar coordinate, also in large scale
PIC simulation as in GTC \cite{Lin1998}. The conventional method to
treat this problem is making a hollow and ignoring a small $r \to 0$
region. While, one can find in \cite{Sadiku2009} other ways of MC
implement transport equations and treat the $r \to 0$  problem
(e.g., L'Hospital's rule $lim_{\rho \to
0}\partial{f}/(\rho\partial{\rho})=\partial{^2}f/\partial{\rho{^2}}$),
which means that this won't be a severe problem in PMC.

\begin{figure}[htp]
\epsfig{file=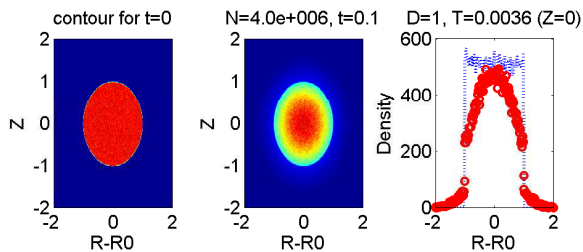,clip=true,width=9cm}
\caption{\label{fig3}(Color online) PMC for (Tokamak) edge pedestal
transport.}
\end{figure}

However, in fact, we will find MC can be a natural method to solve
this geometry and complex boundary problems. The only reason why
people using different coordinates is that we can use the symmetry
of the system to make us treat (e.g., analytical calculations) or
understand problem easier. People care but the great nature never
cares the coordinates. The physics law won't change in any
coordinates. So, we can rewrite (\ref{eqn7}) from $(r,\theta)$
coordinate to Cartesian coordinate $(x,y)$  and use
$\sqrt{x^2+y^2}=r=r_b$ as boundary. MC is suitable for any complex
boundaries, and the circle boundary is just one of them. Then, in
use of PMC, people can use Cartesian coordinate always (in fact, as
the circle symmetry case in FIG.\ref{fig2}, we have already given an
example), which will simplify many old complex coordinates (notable:
the magnetic surface coordinate) or boundaries problems a lot. A 2D
implement of PMC for this edge pedestal transport model is shown in
FIG.\ref{fig3}.
In tokamak experiment, one can see lots of figures as
FIG.\ref{fig3}, i.e., the above simple model can model the
experiment at least qualitatively.

\begin{table*}
\caption{\label{tab1}Comparison of FKC, PIC and PMC for kinetic
plasma simulation.}
\begin{ruledtabular}
\begin{tabular}[c]{|c|c|p{50mm}|p{50mm}|}
   ~ & \textbf{FKC} & \textbf{PIC} & \textbf{PMC} \\ \hline
  Approach &  Continuum &  Particle + continuum &  Only particle \\ \hline
  Equations &  Only PDEs &  New method for eqn. of motion, Field eqn. brown from FKC
 &  Eqn. of motion brown from PIC,
New method for field equation \\ \hline
  Good &  Accurate &  Efficient &  Easy for complex problems,
Efficient for parallel, Easy for coding \\ \hline
  Bad &  Numerical unstability and dissipation & Noise &  Crude, with error $\propto 1/\sqrt{N}$  \\ \hline
  Philosophy &  More mathematical &  Half math, half first principle &
  More first principle \\
\end{tabular}
\end{ruledtabular}
\end{table*}

\textit{Kinetic 1D electrostatic (ES1D) problem}--For ES1D, the
system is described by kinetic equation
\begin{equation}\label{eqn8}
{{\partial f} \over {\partial t}} + v{{\partial f} \over {\partial
x}} + {{qE} \over m}{{\partial f} \over {\partial v}} = C(f).
\end{equation}
where,  $q$ is the charge, $m$  is mass,  $E=-\nabla{\phi}$ is the
electrical field, $C(f)$ is collision term. The field equation is
(\ref{eqn5}), with $\rho=\int{(f_i-f_e)dv}$ the charge density. All
the variables has normalized in units of $e\phi /T_e$,
$\omega_{pe}^{-1}$, $\lambda_{De}$. For 1D, (\ref{eqn5}) can
rewrite to a more simple ODE form
\begin{equation}\label{eqn9}
dE/dx = \rho.
\end{equation}
Firstly, we can easily find (\ref{eqn8}) has the form of
(\ref{eqn1}), gives
\begin{equation}\label{eqn10}
d{x_i}= {v_i}dt, ~d{v_i} = (qE/m)dt,
\end{equation}
which is totally the same as in PIC. Eq.(\ref{eqn10}) tells us that
more dimensions won't bring significant new time cost (note: higher
dimensions we may need more particles for accuracy, which will bring
some new time cost), while (\ref{eqn8}) will if we use DD. If the
collision term has the form
$C(f)=\alpha{f}+\beta{\partial{f}/\partial{v}}+\gamma{\partial{^2}{f}/\partial{v^2}}$
, the kinetic equation is still nothing else but (\ref{eqn1}). We
see here, for PIC simulation of collision, we need an extra MC step,
while in PMC method, we can write this step to equation of motion
(\ref{eqn10}) directly.

Here, we assume ions immobile for simplify and treat the
collisionless (Vlasov) case, i.e., $C(f)=0$ , because that which is
easy for benchmark.

For $f(x,v,t)$ , $x$ and $v$ can be anywhere. For $E(x)$  or
$\phi(x)$ , the $x$ can also be anywhere. But, (\ref{eqn8}) and
(\ref{eqn9}) are coupled. We should find out a way to connect the
position $x$ in $f$ and $E$. A naturally thought is calculating the
$E(x)$ directly using MC for every $f$-particle one by one, but
which is too much time consumed, especially when $f$-particles
($N_f$) and $\phi$-particles ($N_{\phi}$) are very large. An adjust
way is using grids and then interpolating, which can reduce the time
cost by an order of $O(N_{\phi})$. We find this is just what PIC
does, i.e., we re-deduce all the key steps of PIC method by PMC! The
interpolating method is easy and has checked by PIC, however, one
may also find other ways to implement this PMC step if can make sure
they work.

Using ODE (\ref{eqn9}) as field equation, for MC, which is just an
integral from one side to another side, and we use random particles
to do this, the errors of each step will cancel and average is zero.
While, for DD (e.g., Euler, Runge-Kutta or else), the integral is
summation and errors will accumulate. Using Poisson equation, the MC
method is provided at before, an extra step is needed to calculate
$E$ from $\phi$ by calculating gradient from the MC
$\phi$-particles. Periodic boundary for the MC particles of $f$ is
easy to implement, which is the same as PIC. We comment here, for
$\phi$, the periodic boundary means $\phi(0)=\phi(L)$, but when we
see $E$, which is not $E(0)=E(L)$ but average $<E(x)>=0$
(\cite{Birdsall1991} has known this). In fact, this implies that, we
have added two charged slabs (not real 'particles', because this is
1D, or called 'markers' in PIC) at the boundary to cancel the extra
E field! Few ES1D PIC researchers have seen this. Thinking in PMC
way will bring us many new insights or re-thoughts.

For this ES1D problem, we can find, the conventional PIC treatment
and new PMC treatment can be very similar. The implements can be
close, but the concepts are totally different. PIC is fixed and
nonadjustable. PMC provides more possibilities and can bring us new
ways.

\begin{figure}[htp]
\epsfig{file=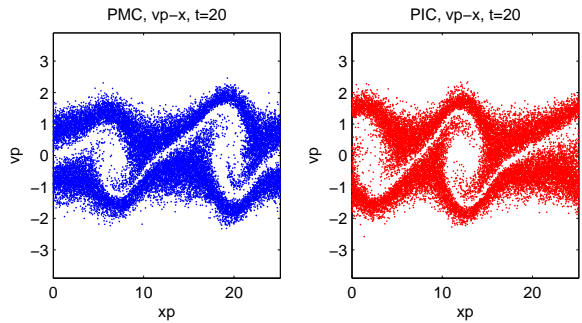,clip=true,width=9cm}
\caption{\label{fig4}(Color online) Compare with PMC and PIC for
2-stream instability, phase space plotting.}
\end{figure}

Using PMC to solving (\ref{eqn8}) and (\ref{eqn9}), a result is show
in FIG. ~\ref{fig4}.
Note that, due to noises, the phase space plotting will change for
both PIC and PMC, even though under the same parameters (e.g., keep
source code unchanged), especially in nonlinear stages. The above
figure is a select from typical runs. The main feature of phase
space holes is clearly in both PMC and PIC results. PMC is slightly
slower than PIC in this case.

\textbf{\textit{Summary and Comments}}--The above two paradigms have
shown many new advantages. In fact, PMC is more powerful when
problems are complicated. For example, many large scaled codes are
possible be completely rewritten via this new method, an example is
BOUT (\cite{Xu2001}, gyrofluid). For large scaled gyro-kinetic, this
method is also possible. Some complicated paradigms are beyond this
short letter.

A comparison of FKC, PIC and PMC is listed in TABEL. \ref{tab1},
which can also shows that why we can call PMC be a third method for
plasma simulation.

At present, many tools of PMC are brown from other fields. But the
concept change is the key important thing. The unified viewpoint by
combining everything in one provides us a totally new picture. In
this new angle, we can see difficult old problems naturally and
problems become simple. We can not only do new problems but also
provide new insights or more efficient methods for old problems. For
kinetic problems, PMC can be seen as advanced PIC; for fluid
problems and complex procedures, PMC is new. Whatever, people may
just concern that if an approach is useful and works well. PMC
matches this. PMC is not a single fixed approach as PIC, but a
combine of varies of old and new MC methods, flexible.

As far as the author (HSX) knows, Max-Planck-Institut f\"ur
Plasmaphysik is developing MC methods for Stellarator edge transport
physics, while it seems \cite{Sengil2012} has already or almost
implemented a fully PMC ES2D code though they didn't know this
concept when they did that. New applications for resistive MHD
tearing mode, Hasegawa-Mima equation, Grad-Shafranov equation and
many other problems are on the way by the author or others. This
letter has given/built the framework of PMC, which is enough for
many practical applications. But, there may are still many new
intelligent works need do for some special types of cases. Some
unexpected problems may also meet when using to more complicated
examples. The author does not know that yet. If also implemented
well for the above untested several examples, this new approach is
possible completely substitute the conventional two methods, FKC and
PIC, and also suitable for solving many old unsolved or hard to
solve problems, then bring plasma simulation to a new era.

\textbf{\textit{Acknowledgement}}--This work is also inspired by
Zhi-chen FENG (IFTS-ZJU), who once (2010) proposed a new method for
PIC, though which performances not well in practical application
yet.


\end{document}